\documentclass{KapProc} 

\usepackage{graphicx}
\kluwerbib

\def \aj {AJ}
\def \aap {A\&A}
\def \aaps {A\&A Suppl.}
\def \araa {ARA\&A}
\def \apj {ApJ}
\def \apjl {ApJL}
\def \nat {Nature}
\def \mnras {MNRAS}
\def \pasp {PASP}

\begin{document}

\articletitle[Supernova types and rates]
{SUPERNOVA TYPES AND RATES \thanks{Most of the material presented in
this paper has been retrieved from the Padova Archive of SN
observations. This is the product of a long standing effort for the
monitoring of SNe conducted using ESO-La Silla and Asiago telescopes
(\cite{kp}).}}

\author{Enrico Cappellaro}
\email{cappellaro@pd.astro.it}
\author{Massimo Turatto}
\email{turatto@pd.astro.it}

\affil{Osservatorio Astronomico di Padova\\ vicolo dell'Osservatorio, 5\\
I-35122, Padova (Italy)}


\begin{abstract}
We review the basic properties of the different supernova types
identified in the current taxonomy, with emphasis on the more recent
developments. To help orienting in the variegate zoo, the optical
photometric and spectroscopic properties of the different supernova
types are presented in a number of summary figures.

We also report the latest estimates of the supernova rates and stress
the need for a dedicated effort to measure SN rates at high redshift.

\end{abstract}


\section{Introduction}

Supernovae (SNe) are at the confluence of many different streams of
astronomical researches. As the final episode of the life of many
kinds of stars, SNe allow crucial tests of stellar evolution
theories. Yet,  in many cases we are still
entangled with the basic problem of finding a viable progenitor
scenario for each SN type.  During and immediately after the
explosions, there are a variety of fundamental physical mechanisms
which can be probed, such as neutrino and gravitational wave
emissions, flame propagation and explosive nucleosynthesis,
radioactive decays and shocks with circumstellar matter. Relics of the
explosions are collapsed remnants, neutron stars or black holes, and
expanding gas clouds which heat and pollute the interstellar
medium.  Indeed, SNe are a main agent in the chemical
evolution of galaxies.

Moreover, because of their huge luminosity and the fact that they can
be accurately calibrated, type Ia SNe are the best distance indicators
for measuring the geometry of the Universe.

For all these reasons, in the last decades there has been an increased
interest in SN studies and a renewed effort in SN searches.  As a
consequence, the number of SN discoveries has increased from about
20/yr in the early '80, to about 200/yr in the last few years raising
the number of SNe discovered to date to $\sim 1800$ (\cite{cat};\\
\verb|http://merlino.pd.astro.it/~supern/|
 for a frequently updated
version). It is not surprising that the enlarged sample combined with
the better quality of the observations obtained by the new generation
instruments is deeply changing our understanding of the SN
phenomenon.

\begin{figure}[t]
\resizebox{\hsize}{!}{\includegraphics*{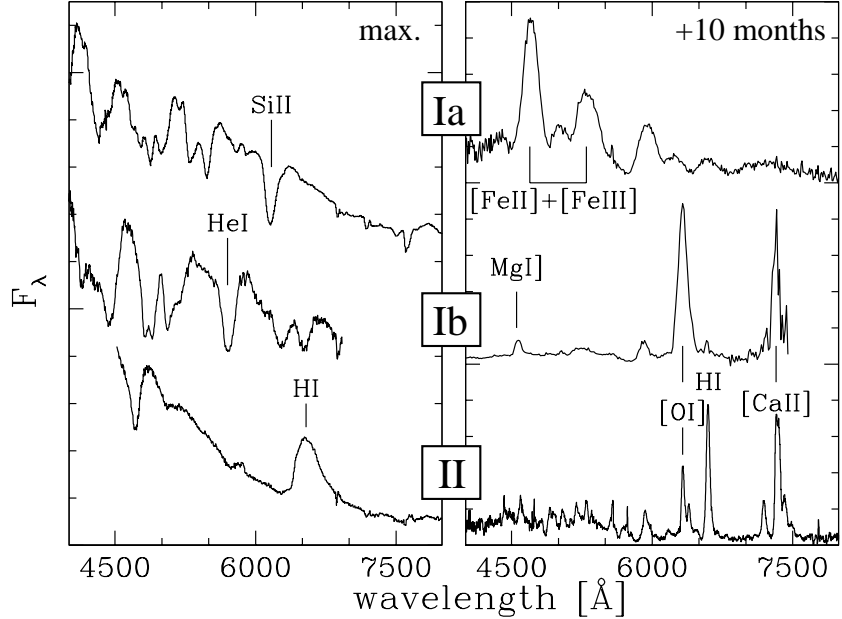}}
\caption{Basic SN types spectra.  A SN which near maximum (left panel) 
shows clear signature of H$\alpha$ is defined as type II, if it shows
a strong SiII absorption at about 6150 \AA\/ is a type Ia,
otherwise it is of type Ib/c (in the figure we show a SNIb 
characterized by strong
He lines). Ten months later
(right panel) SN~Ia show strong emissions of [FeII] and [FeIII],
SN~Ib/c are dominated by [CaII] and [OI]. These same lines and
strong H$\alpha$ emission are typical of SNII.}\label{basic}
\end{figure}

The aim of this paper is to give a snapshot of the SN zoo including
the latest acquisitions. Other information can be found in
Filippenko (1997), Wheeler and Benetti (2000) and reference
therein. We will also present the latest estimates of the SN rates
which are fundamental to link the evolution of stars and stellar
systems.

\section{Basic SN types}

The affair of SN classification began with Minkowskii (1941)
who noticed that there are at least two different kinds of SNe, 
showing (type II) and not showing (type I) H in their spectra. 

This subdivision was maintained until, in the '80s, it was recognized
that a number of peculiar SN~I missing in the spectrum near maximum the
typical feature at $\sim 6510$ \AA, had a completely
different nebular spectrum dominated by forbidden Ca and O emissions
instead of Fe lines (\cite{gaskell}). Examining more carefully the
near maximum spectra it was noticed that these SNe come in at least
two flavors: those showing strong He lines were labeled Ib, the
others Ic. We will argue later however that this subdivision is at a
lower hierarchical level.  The first branch of the SN classification
is illustrated in Fig.~\ref{basic}.

\begin{figure}[t]
\resizebox{\hsize}{!}{\includegraphics*{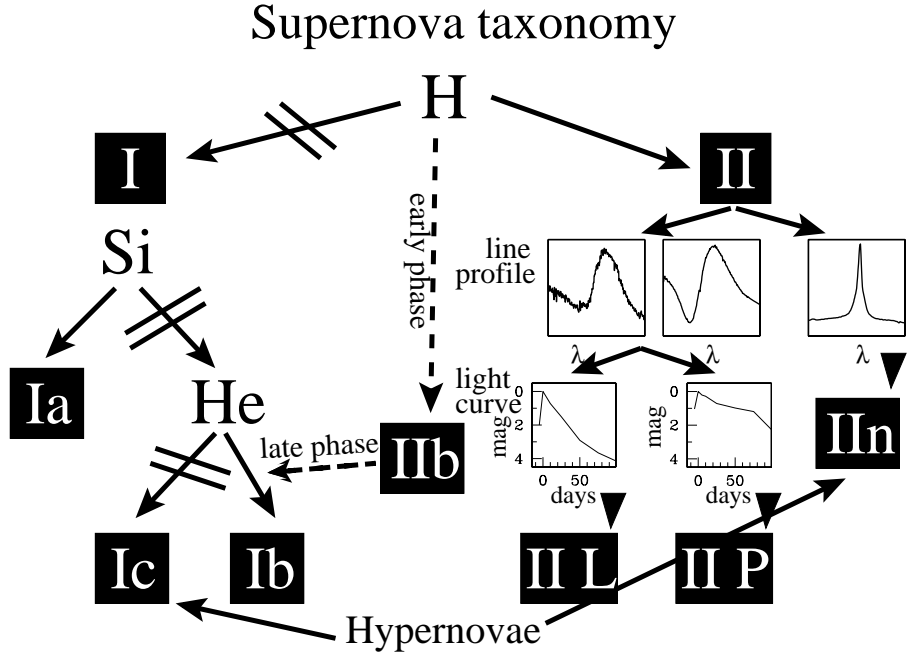}}
\caption{The detailed classification of SNe requires not only the 
identification of 
specific features in the early spectra, but also the analysis
of the
line profiles, luminosity and spectral evolutions}\label{taxo}
\end{figure}

In general, in the early phases the optical depth of the ejecta
remains high and the emergent spectrum only probes the outermost
layers. Therefore the early spectrum is most sensitive to variations
in the density and composition of the progenitor envelope.  With time
the density and temperature decrease, the photosphere recedes in mass
coordinates and eventually the ejecta becomes optically thin. At this
point the innermost regions becomes exposed and the yields of the
explosions can be probed. The fact that the nebular spectrum of SN~Ib/c
is similar to that of SN~II, but for the H emission, prompts for a
similar explosion mechanism.  Instead, the weakness of
intermediate element emissions, in particular O, in the nebular
spectrum of SN~Ia excludes that the progenitors of SN~Ia are massive
stars.

The SN taxonomy which originally was based on spectra near maximum may
result somewhat confusing when one try to build a coherent progenitor
scenario: indeed a more physical classification could be based on
the distinction between SNe arising from the collapse of massive stars (SNII
and SNIb/c) and SNe due to thermonuclear explosions of low mass stars
(SN~Ia).  In the following we will give a brief description of the
characteristics of each SN type describing the  detailed taxonomy
illustrated in Fig.~\ref{taxo}.

\begin{figure}[t]
\centering
\resizebox{11cm}{!}{\includegraphics*{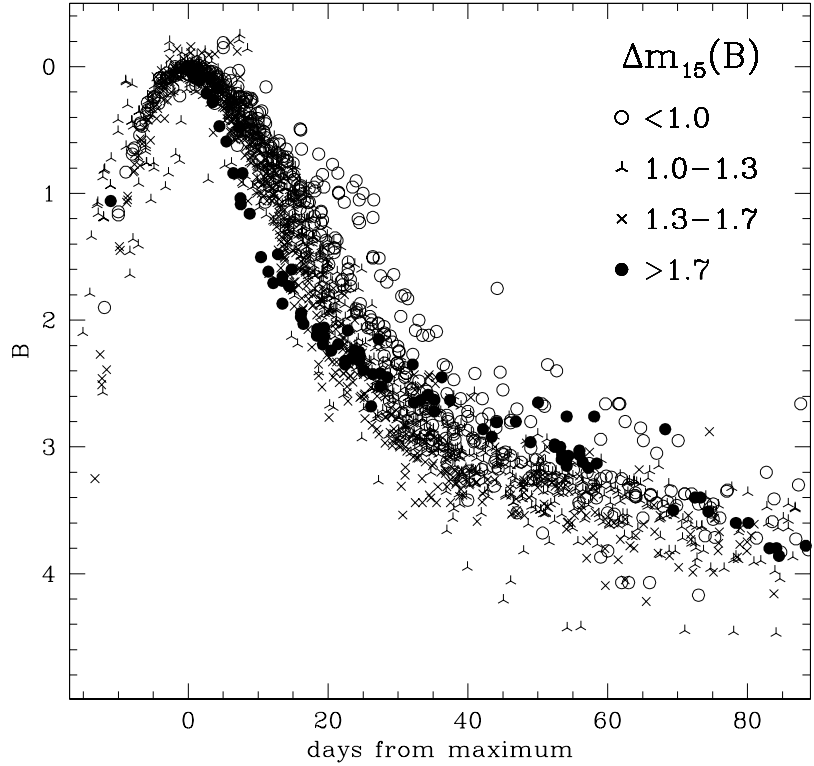}}
\caption{Light curves of 87 SN~Ia relative to the phase and magnitude 
of maximum. The symbols identify objects with different $\Delta
m_{15}(B)$, the magnitude decline in the early 15 days (Altavilla
2000).}\label{lcia}
\end{figure}

\section{Type Ia SNe} 

\begin{figure}[t]
\resizebox{\hsize}{!}{\includegraphics*{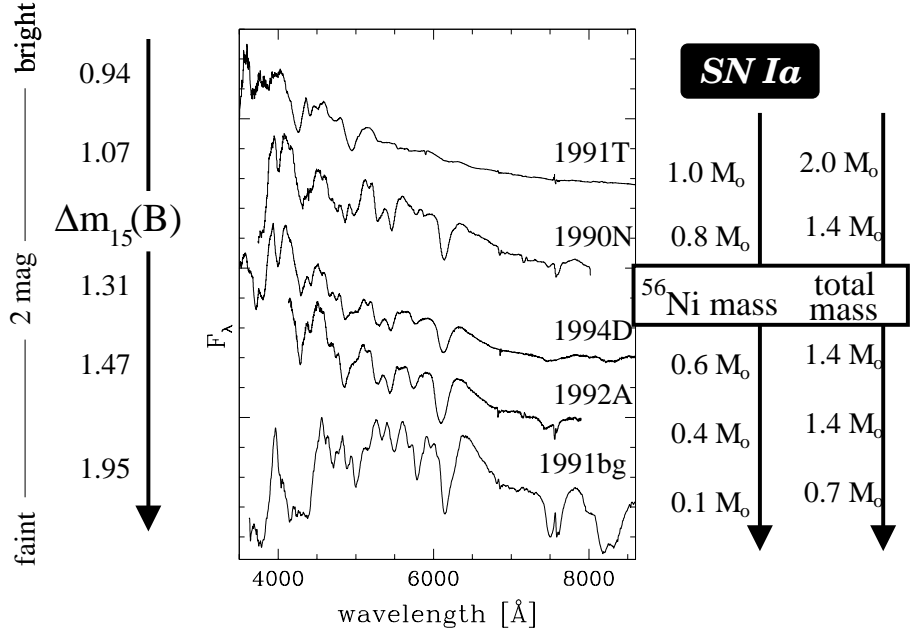}}
\caption{Comparison of the maximum light spectra of SN~Ia with different 
decline rates (left), hence luminosities.  On the right the ranges in
$^{56}$Ni and total masses suggested by light curve and spectral
synthesis models are indicated.}\label{specia}
\end{figure}

SN~Ia are quite homogeneous events 
with similar luminosity and spectral evolution.  Indeed until the
early '90s, it was commonly accepted that
SN~Ia were identical and that
observed differences were mainly due to observational errors (e.g
\cite{bruno}). With the improvement of the signal to noise of the
observations it was definitely demonstrated that differences do
exist. In particular, though the light curve shapes remain similar,
the absolute luminosities and the decline rates are different
(Fig.~\ref{lcia}). Actually, a relation has been found between the
rate of the luminosity decline and the absolute magnitudes at maximum,
with fast declining SNe being fainter (and redder). Because this is
crucial for recovering SN~Ia as distance indicators, a large efforts
is being devoted to the accurate calibration of the different
expressions of this relation (\cite{phil},
\cite{riess}, \cite{perlm}).  To some extent, the spread in luminosity
correlates with variations in the spectral properties
(\cite{nugent}). As shown in Fig.~\ref{specia}, very bright or very
faint SNe show distinctive spectral features, whereas the
differences among ``normal'' SN~Ia stand out only after 
a careful analysis.

It is generally agreed that the difference in luminosity call for one
order of magnitude variance in the mass of the radioactive $^{56}$Ni
which is powering the early light curve. Controversial is the claim
that light curves and spectral synthesis models require a factor two in
the total progenitor mass (\cite{cap97}).  The standard scenario for
the SN~Ia explosion consists of a binary system where one of the star
is a white dwarf (WD) which accretes matter for the secondary star,
reaches the Chandrasekhar limit, and undergoes a disruptive,  thermonuclear
explosion  (\cite{nomoto},\cite{woosley}).  
It is still unclear whether
it is possible to reproduce the observed variance of SN~Ia within the
standard scenario (\cite{mazzali}).

\begin{figure}[t]
\centering
\resizebox{8cm}{!}{\includegraphics*{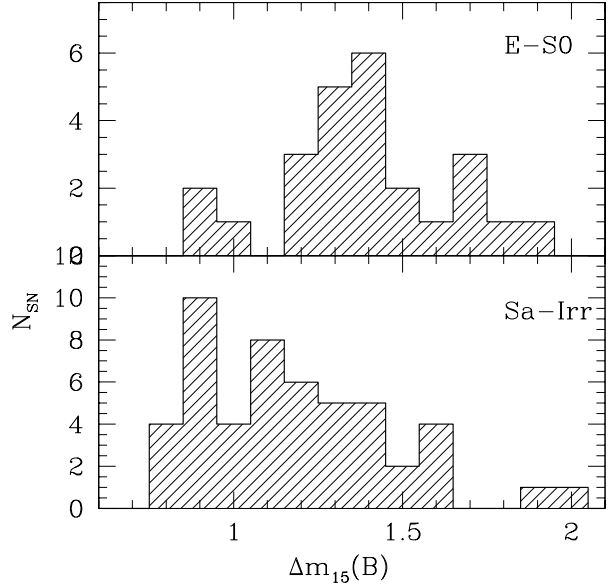}}
\caption{Histogram of $\Delta m_{15}(B)$ for SN~Ia in early type
galaxies (upper panel) and in late type galaxies (lower panel).  A
Kolmogorov-Smirnov test shows that the probability that the
two distributions derive from the same population is only 0.004
(Altavilla 2000).}\label{dm15ist}
\end{figure}

An interesting finding in this respect is that on average SN~Ia in
early type galaxies show a faster decline rate (hence are less
luminous) than SN~Ia in late type galaxies (Fig.~\ref{dm15ist},
cf. \cite{vdb},\cite{hamuy}).  This suggests that (some of) the
variance of SN~Ia is related to different ages of the progenitor
population. Until this is not well understood, a severe caveat
remains on the use of SN~Ia as cosmological distance indicators.

\section{Type II SNe}

\begin{figure}[t]
\centering
\resizebox{11cm}{!}{\includegraphics*{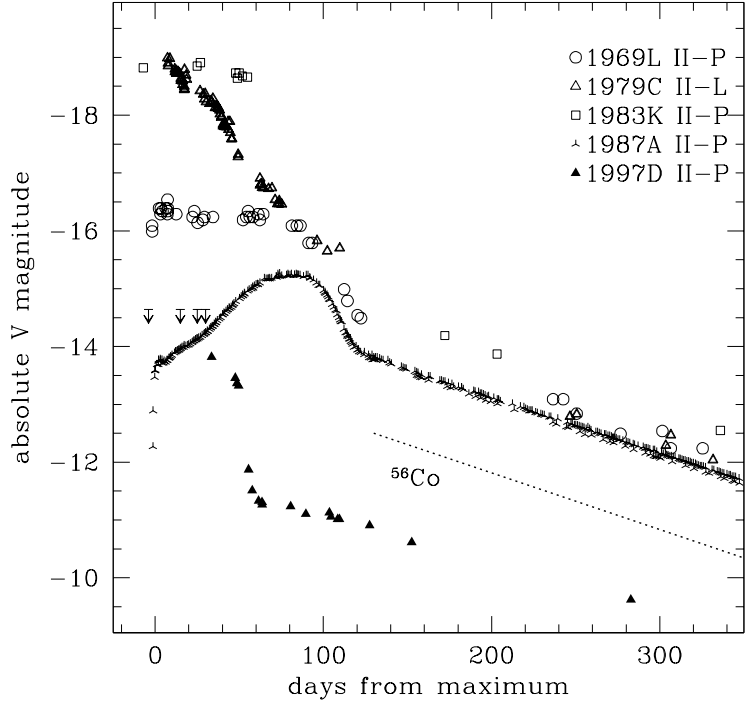}}
\caption{Representative light curves of SN~II. The dotted line is the 
expected luminosity decline rate when the light curve is powered  
by the decay of $^{56}$Co.}\label{iilc}
\end{figure}

One might say that the only thing in
common for SN~II is the presence of H lines in the spectra. Indeed
the maximum luminosities range over two orders of magnitudes and the
spectral energy distributions and line profiles also shows dramatic
differences. Historically, the first attempt to sort out some order
was based on the light curve shapes (\cite{barbon}). In some cases, SN~II
after maximum remain on a plateau of almost constant luminosity for
periods up to 2-3 months (II-P), in other cases they decline more or
less linearly (II-L). The two
classes are not separated and intermediate cases do exist (\cite{cloc}). 
An extreme
case of SN~II-P is usually considered SN~1987A, the best observed SN
ever, which after maximum showed a steadily increase of magnitude,
lasting abound 3 months, until eventually the decline began (Fig.~\ref{iilc}).

In general, it is assumed that SNII-L are brighter than SN~IIP, but
there are many exceptions (\cite{patat}), e.g. SN~1983K the
bright II- P shown in Fig.~\ref{iilc}. At late time ($>200$ d), many
SN~II settle on a linear decline rate of about 1 mag/100d. This is
powered by the second branch of the radioactive decay chain
$^{56}$Ni--$^{56}$Co--$^{56}$Fe.  The fact that, despite the enormous
difference at maximum, SN~II at late time converge to a similar
luminosity was taken as an indication that the $^{56}$Ni mass produced
in SN~II was very similar, namely $\sim 0.1$ M$_\odot$
(\cite{turatto}). Actually, even in this respect there may be
significant variations (\cite{schmidt}).  An extreme case is that of
SN~1997D (Fig.~\ref{iilc}) which produced only 0.002 M$_\odot$ of
$^{56}$Ni and showed extremely low expansion velocities (\cite{97d}).

\begin{figure}[t]
\resizebox{\hsize}{!}{\includegraphics*{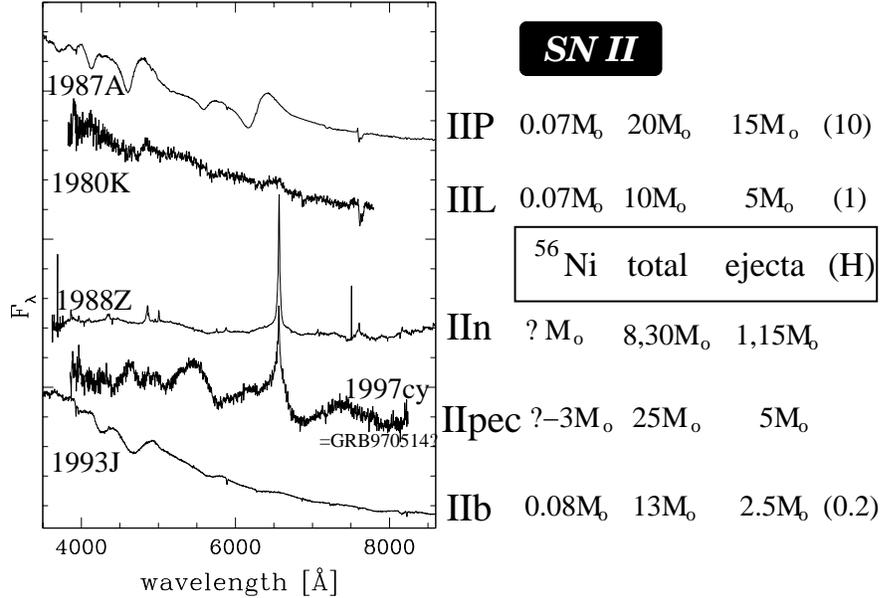}}
\caption{Representative spectra of SN~II. On the right
we report for each object the best estimate of the $^{56}$Ni, total
and ejecta masses (in parenthesis is the H mass in the
ejecta). }\label{specii}
\end{figure}

It is generally accepted that SN~II result from the core collapse of
stars in the range of 10-30 M$_\odot$ (\cite{woosley87a}). Because of
various mechanisms (eg. difference in metallicities, interaction in
binary system) the size and the mass of the H envelope at the time of
explosion can be very different even for progenitors of the same
initial mass.  In a typical SN~IIP the H envelope mass is $\sim 5-10$
M$_\odot$ and the radius is $\sim 10^{15}$ cm. The shock wave
originated by the explosion rapidly reaches the surface completely
ionizing the H. When the ejecta expands the energy from the
recombination of H sustains the photosphere at an almost constant
radius and temperature.  The length of the plateau depends on the
envelope mass and when this gets very small ($<1-2$ M$_\odot$) the
decline is faster (SN~IIL).  The different line profiles of II-P and
II-L, with II-L missing the P-Cygni absorption of the Balmer lines
(Fig.~\ref{specii}) is also due to their reduced envelope masses.

Instead the early luminosity depends on the radius of the progenitor
star at the time of explosion. If the progenitor was compact,
$10^{12}$ cm in the case of SN~1987A, much of the available energy is
dissipated for the expansion and until the radioactive decay input
becomes dominant, the luminosity remains lower than in the case of an
extended progenitor.

In most cases, the remnant of the explosion is expected to be a neutron
stars, but it has been argued that the unusual properties of
SN~1997D may be understood if the collapsed remnant is a black hole
(\cite{97d}).

\begin{figure}[t]
\centering
\resizebox{11cm}{!}{\includegraphics*{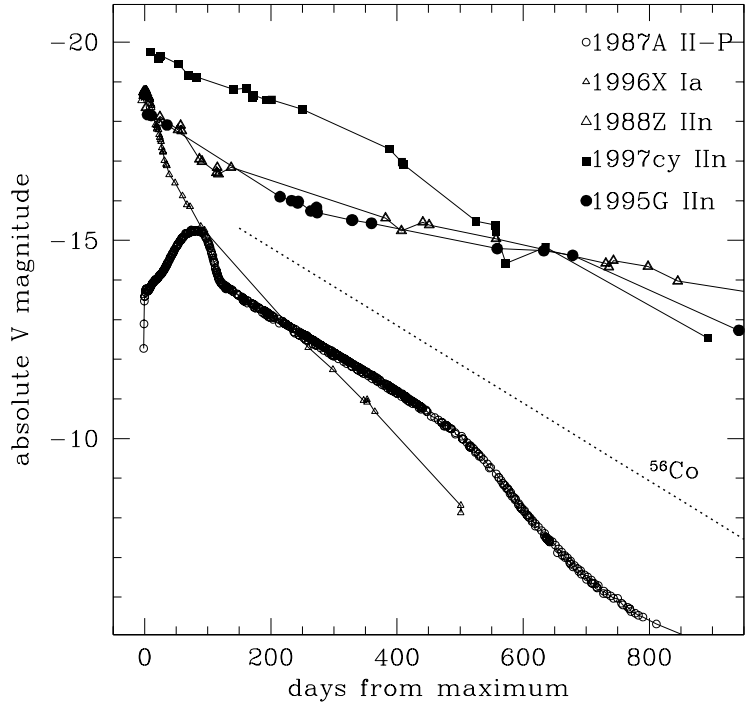}}
\caption{Representative light curves of SN~IIn are compared with that
of a SN~II-P and of a SN~Ia. The dotted line is as in Fig.~\ref{iilc}}
\label{iinlc}
\end{figure}

In general, because of the very high expansion velocity, the emission
lines in the SN spectra are very broad but for possible contamination
of interstellar gas emissions. A significant fraction of SN~II
($10-15$ \% of the SN~II currently discovered) show narrow components
on top of broader emissions (Fig.~\ref {specii}). These SNe are
labeled IIn, where ``n'' stands for narrow--line (\cite{schlegel}).
It turns out that in many cases SN~IIn are very bright and their
luminosity evolution is much slower than other SN~II
(Fig.~\ref{iinlc}).  Therefore, an additional source of energy
powering the light curve in addition to the radioactive decay is
required.  It is currently believed that the kinetic energy of the
ejecta is converted into radiation when the ejecta shock a dense
circumstellar medium (CSM) (\cite{88z}).  This dense CSM is most
likely a relic of a strong mass loss episodes occurring shortly before
explosion. Differences in density and distribution of the CSM might
explain the large variance of luminosity evolution between different
SN~IIn.

\begin{figure}[t]
\centering
\resizebox{11cm}{!}{\includegraphics*{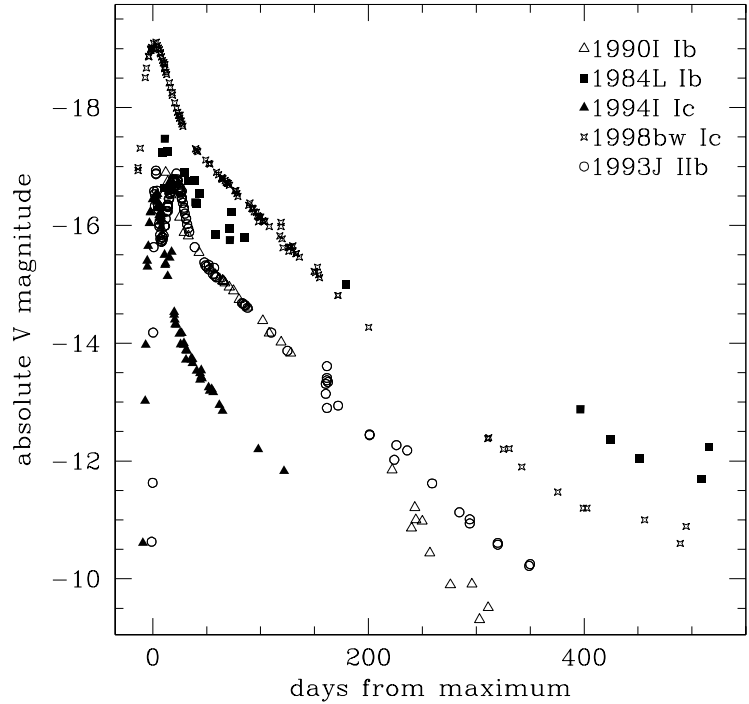}}
\caption{Representative light curves of SN~Ib/c. }\label{ibclc}
\end{figure}

A most interesting case is that of SN~1997cy for its possible
association with GRB970514. Because of the emission line profile
(Fig.~\ref{specii}), we believe that also in this case the CSM-ejecta
interaction is powering the light curve (Fig.\ref{iinlc}) and a huge Ni
mass is not required (\cite{97cy}).

\section{Type Ib/c SNe}

The explosion of SN~1993J in M81 was the missing link between SNe of
type II and Ib/c. The spectrum of this SN evolved from that of a
rather normal type II to that of a type Ib in few weeks. The object is
therefore classified as the prototype of the intermediate class of SN
IIb.  Its observed properties have been explained with the explosion
of a massive progenitor with a very small H envelope (0.2 M$_\odot$).
It is believed that if the exploding star was left without the H
envelope, the SN would appear as a typical SNIb. If even the He shell
was removed the result will closely match the observational properties
of type Ic.  Other cases of SN~Ic in which is detected a small amounts
of H in the ejecta are reported in the literature (cf. \cite{87m};
2000H, \cite{2000h}). There are also intermediate cases between type
Ib and Ic in which residual amounts of He have been detected
(eg. \cite{94i}).  These considerations have motivated the
introduction of a unifying scenario where the sequence Ic-Ib-IIb-IIL
has mass loss as the driving parameter (\cite{iilic}). At the moment
the mechanism for this huge mass loss has not yet been identified,
although interaction in close binary systems is the first candidate.

\begin{figure}[t]
\resizebox{\hsize}{!}{\includegraphics*{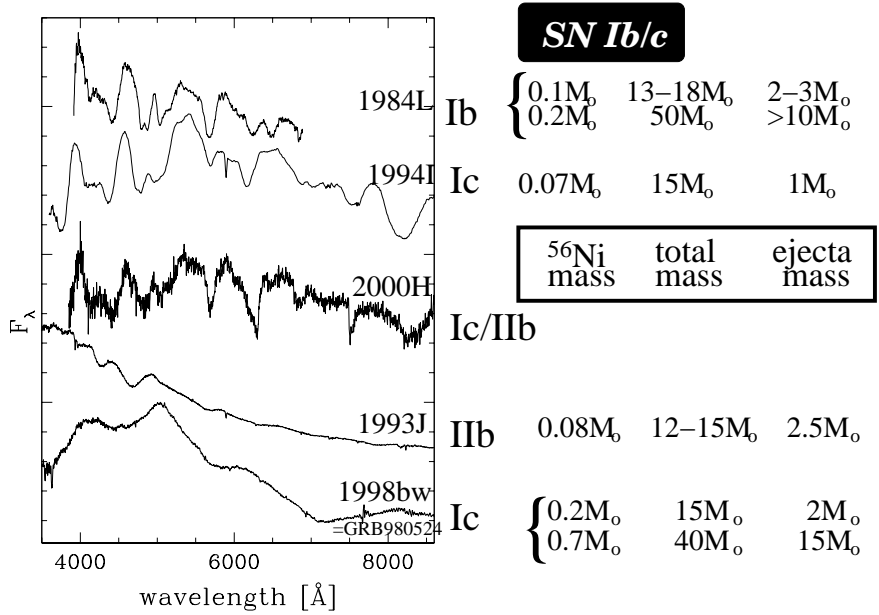}}
\caption{Representative spectra of SNIb/c. On the right
the best estimates of the $^{56}$Ni, total
and ejecta masses are reported . In some case different modeling produces 
significantly different results.}\label{specibc}
\end{figure}

One additional complication is that there are SN~Ib/c in which the
late light curves seem to track the radioactive decay of $^{56}$Co
(eg. SN~1984L) and other showing a luminosity decline similar or even
faster than that of SN~Ia (Fig.~\ref{ibclc}).  Because in all cases it
is believed that radioactive $^{56}$Co is present, one must admit that
the trapping of the $\gamma$-rays from $^{56}$Co is different which in
turn call for a much smaller ejecta mass in fast declining SNIb/c than
in the slow declining ones. A summary of the $^{56}$Ni and ejecta
masses for some SNIb/c is reported in Fig.~\ref{specibc}

More recently two unusual SNe were discovered (SN 1997ef and 1998bw)
that, although formally classified as type Ic because they miss the
6150 \AA\/ dip of SNIa, the H line of type II and the He lines of type
Ib, show indeed a very peculiar spectrum (Fig.~\ref{specibc}). This is
the result of an extremely high expansion velocities and in turn, of
high explosion energy (Iwamoto et al. 1998, 2000).  Most interesting
is the case of SN~1998bw because of his close association with
GRB250498 (\cite{nandobw}) Because of these extreme characteristics
these SNe, along with SN~1997cy mentioned in the previous section, are
often referred as {\bf hypernovae} and may indeed require specific
progenitor scenarios and explosion mechanism.

\section{SN rates}

The most recent estimate of the rates of the different SN types 
has been published in Cappellaro et al. (1999) and are summarized in
Tab.~\ref{rate}.  In average, these values are in fair agreement with
previous estimates (cf. \cite{vdb&t}).

\begin{table}[h]
\caption{The local SN rates. Units are $SNu = {\rm SN}/10^{10} 
{\rm L}_{\odot,{\rm B}}/ 100 {\rm yr}$. h=H$_0$/100.  }\label{rate}
\centering
\begin{tabular}{lcccc}
\sphline
galaxy    & \multicolumn{4}{c}{SN type}\\
\cline{2-5}
type      &     Ia        &      Ib/c    &    II      &  All   \\
\sphline
E-S0      &  $0.32\pm.11~h^{2}$ &    $<0.02~h^{2}$   &   $<0.04~h^{2}$  & $0.32\pm.11~h^{2}$   \\
S0a-Sb    &  $0.32\pm.12~h^{2}$ & $0.20\pm.11~h^{2}$& $0.75\pm.34~h^{2}$& $1.28\pm.37~h^{2}$\\
Sbc-Sd    &  $0.37\pm.14~h^{2}$ & $0.25\pm.12~h^{2}$& $1.53\pm.62~h^{2}$& $2.15\pm.66~h^{2}$ \\
          &               &             &             & \\
All       &  $0.36\pm.11~h^{2}$ & $0.14\pm.07~h^{2}$& $0.71\pm.34~h^{2}$& $1.21\pm.36~h^{2}$\\
\sphline
\end{tabular}
\end{table}

One interesting
result is that the rate of SN~Ia {\bf per unit B luminosity} is
constant from ellipticals to late spirals. Even if one may argue about
what is the stellar population which dominates the blue luminosity in
spirals, a significant fraction must derive from young stars. This
immediately implies that the average age of the SN~Ia progenitors
is shorter in spirals than in  ellipticals (cf. Fig.5).
Also worth noticing is that the SN Ib/c are only 15\% of all core
collapse events.

Based on the values of Tab.~1, we can derive an estimate of
the SN rate in our own Galaxy making the basic assumption that the
Galaxy has an average SN rate for his morphological type (here assumed
Sb-Sbc) and luminosity ($2.3\times10^{10}L_{B,\odot}$).  If we adopt
${\rm H}_0=65\,{\rm km}\,{\rm s}^{-1}\,{\rm Mpc}^{-1}$, we expect
$0.4\pm0.2$ SN~Ia and $1.5\pm1.0$ SNII+Ib/c per century, that is
roughly one SN every 50 years. Formally, this is a factor two lower
than derived from counts of historical SNe and SN remnants
(eg. \cite{vdb&t}, \cite{strom}). When considering
the large uncertainties however, the two values are not in
disagreement.

For what concerns the SN rate in the local Universe, we have now
reached the point where the statistics is  no longer limited  by the SN
discoveries but by the sample of galaxies for which the necessary
information (distance, luminosity, morphological types and
inclination) are available. In other words, to improve significantly
the statistical accuracy of the results, one needs not
only to continue SN searches but also to obtain more complete galaxy
catalogues.  

A major effort is still required, instead, in
``un-biased'' SN search at high redshift. In fact, the
evolution of the SN rate with redshift brings unique information on
the progenitor scenarios, galaxy star formation history and initial
mass function (eg. \cite{yung}, \cite{frans}).

\begin{table}[h]
\caption{Estimate of SN rates at different redshift. $h=H/100$}\label{ratez}
\begin{centering}
\begin{tabular}{lcccl}
\sphline
 $ <z>$      &  Ia                     &   II          & N(SN)     & source \\
           & $h^2$ SNu              & $h^2$ SNu    &           & \\
\sphline  
 0.01      & $0.36\pm0.10$           & $0.85\pm0.41$ &  137     & \cite{rate99} \\
 0.1       & $0.44^{+.0.35}_{-0.21}$ &               &    4     & \cite{hardin} \\
 0.38      & $0.82^{+.0.65}_{-0.45}$ &               &    3      & \cite{pain}               \\
 0.55      & $0.62\pm0.19$           &               &   38      & Pain et al. 2000$^{\#}$   \\
\sphline
\end{tabular}
\end{centering}

\# Not yet published. Reported in \cite{ruizrate}.
\end{table}

Current high-z SN searches are aimed to the discovery of SN~Ia for
their use as distance indicators. By construction, they are not well
suited for estimating SN rates.  This is obvious from Tab.~\ref{ratez}
where we list all observational estimates of SN rates available at
``high'' redshift and compare them with the local estimates. In this
table N(SN) gives the statistics of SNe on which the estimate is
based. The very low statistics and the fact that there is not an
estimate yet for core-collapse SN rate outside the local Universe is a
clear indication of the work to be done.

\end{document}